\documentclass[prc,showpacs,preprintnumbers,amsmath,amssymb,superscriptaddress,floatfix,nofootinbib]{revtex4}
\usepackage{amsmath,amssymb}
\usepackage{epsfig}
\usepackage{color}
\usepackage{bm}
\usepackage{epstopdf}

\makeatletter
\def\slashchar#1{{\mathpalette\c@ncel{#1}}} 
\makeatother

\begin{document}

\title{Neutrino-nucleus CCQE-like scattering}


\author{J. Nieves}
\affiliation{Instituto de F\'\i sica Corpuscular (IFIC), 
Centro Mixto CSIC-Universidad de Valencia, Institutos de
Investigaci\'on de Paterna, Apartado 22085, E-46071, Valencia, Spain}
\author{R. Gran}
\affiliation{Department of Physics, University of Minnesota--Duluth, Duluth, Minnesota 55812, USA} 
\author{I. Ruiz Simo}
\affiliation{Departamento de F\'\i sica At\'omica, Molecular y Nuclear,
and Instituto de F\'\i sica Te\'orica y Computacional Carlos I, Universidad de Granada, Granada 18071, Spain}
\author{F S\'anchez}
\affiliation{Institut de Fisica d'Altes Energies (IFAE), Bellaterra,
  Barcelona, Spain} 
\author{M. J. Vicente Vacas}
\affiliation{Instituto de F\'\i sica Corpuscular (IFIC), 
Centro Mixto CSIC-Universidad de Valencia, Institutos de
Investigaci\'on de Paterna, Apartado 22085, E-46071, Valencia, Spain}

\begin{abstract}
RPA correlations, spectral function and 2p2h
(multi-nucleon) effects on charged-current neutrino-nucleus
reactions without emitted pions are discussed. We pay attention to the influence of RPA and multi-nucleon mechanisms on  
the MiniBooNE and MINERvA flux folded  differential cross sections, the MiniBooNE flux unfolded total cross section  and  the neutrino energy
reconstruction.
\end{abstract}
\maketitle
\section{Introduction}

The origin of the so called MiniBooNE charged-current quasi-elastic (CCQE) puzzle has been extensively debated (see for instance Refs.~\cite{Morfin:2012kn,Alvarez-Ruso:2014bla}) since this collaboration presented in 2009 a new 
CCQE cross section measurement~\cite{AguilarArevalo:2010zc} using a
high-statistics sample of $\nu_\mu$ interactions on $^{12}$C. The experiment accounted for events with no pions in the final state, but Monte Carlo correcting for those cases where  CC
pion production was followed by pion absorption. It was customary to take for granted that most of those events could be attributed to the
QE scattering of the weak probe on a nucleon, and thus the initial neutrino  energy could be approximately
determined from the energy and angle of the final lepton assuming QE kinematics. In what follows, we will refer as QE-like to this data sample.  However, the size of the QE-like 
cross section was found to be unexpectedly large, and within 
the relativistic global Fermi gas model employed in the analysis, a difficult to accept\footnote{The value of $M_A$ extracted from early CCQE measurements on deuterium and, to a lesser extent,
hydrogen targets is $M_A = 1.016 \pm 0.026$ GeV \cite{Bodek:2007ym}, which is in excellent agreement with the pion
electro-production result, $M_A = 1.014 \pm 0.016$ GeV, obtained from the nucleon axial radius~\cite{Alvarez-Ruso:2014bla, Bernard:1992ys}. Furthermore,  NOMAD also reported in 2008 a small value of
$M_A = 1.05 \pm 0.02~({\rm stat})~ \pm 0.06~ {\rm (syst)}$ GeV \cite{Lyubushkin:2008pe}. Nevertheless, we will make a further comment on the NOMAD result below.}  large nucleon axial mass of
$M_A = 1.35 \pm 0.17$ GeV was needed to describe the data. Moreover, the results of Ref.~\cite{Benhar:2010nx}, based on the impulse approximation
scheme and a state-of-the-art model of the nuclear spectral functions, suggested that the electron cross
section and the MiniBooNE flux averaged neutrino cross sections, corresponding to the same target and comparable
kinematical conditions, could not be described within the same theoretical approach using the value of the
nucleon axial mass obtained from deuterium measurements.

A  natural solution to this puzzle comes from the incorporation of RPA and multinucleon nuclear effects. Indeed, the QE-like
sample includes also multinucleon events where the gauge boson is absorbed by two interacting nucleons (in the many body language, this
amounts to the excitation of a 2p2h nuclear component). Up to re-scattering
processes which could eventually produce secondary pions, 2p2h events will give rise to only one muon to be detected.
Thus, they could be experimentally misidentified as QE events.

The importance of 2p2h effects for QE-like scattering was first explored in Refs.~\cite{Martini:2009uj,Martini:2010ex}  and
later in Refs.~\cite{Amaro:2010sd,Nieves:2011pp,Nieves:2011yp}. Some of these more complete models, that also account for long range RPA corrections,  
were found to describe well even the MiniBooNE double differential (2D) cross section while using a standard value, of the order of 1 GeV for 
$M_A$~\cite{Nieves:2011yp,Martini:2011wp}\footnote{Indeed the microscopic model of Ref.~\cite{Nieves:2011pp}, used in \cite{Nieves:2011yp} to analyze the MiniBooNE 2D neutrino data, provides also 
a fair description~\cite{Nieves:2013fr} of the later MiniBooNE 2D cross section measurements for antineutrinos~\cite{AguilarArevalo:2013hm}. 2D antineutrino data were also well reproduced~\cite{Martini:2013sha} 
within the model of Refs.~\cite{Martini:2009uj,Martini:2010ex}.}.

\begin{figure}[tb] 
\begin{center}
\includegraphics[width=0.25\textwidth]{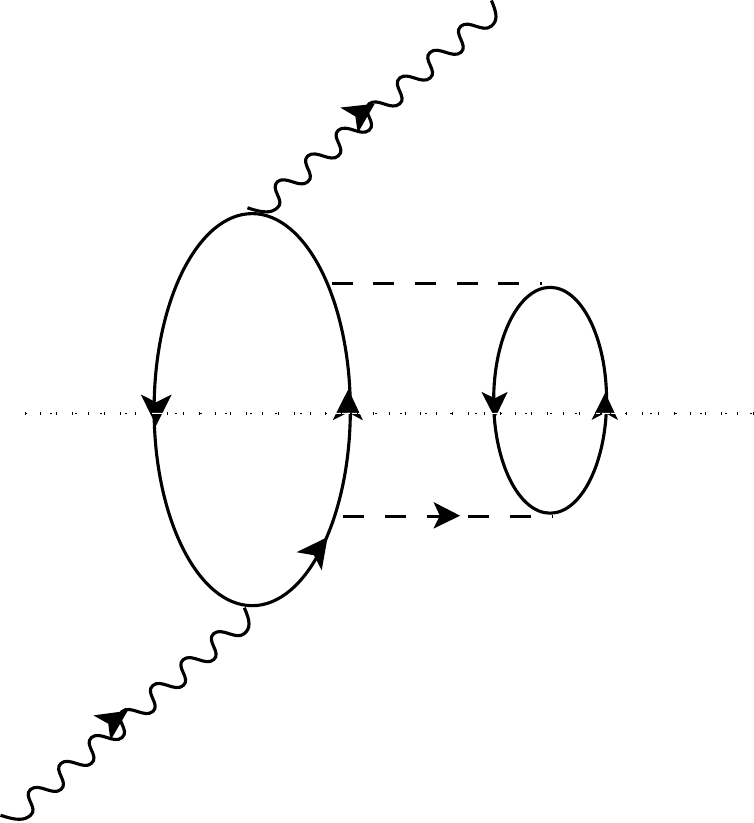}\\
\includegraphics[width=0.2\textwidth]{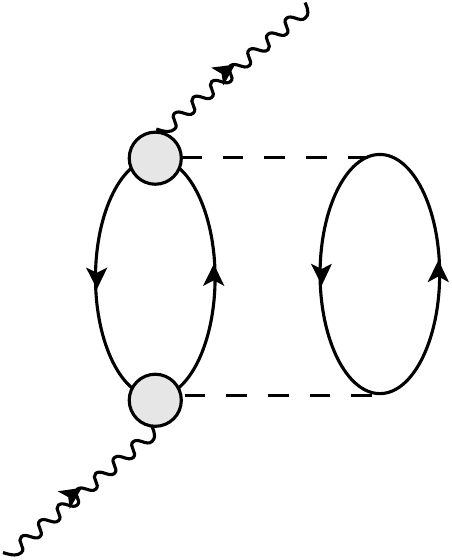}
\caption{ Top: $W$ self-energy diagram obtained from the genuine QE term by dressing up the nucleon propagator of the particle state. Bottom: Generic 2p2h contribution to the  $W$ self-energy, where the circle stands for
a full elementary model for the $WN \to \pi N$ process.}\label{fig:diag}
\end{center}
\end{figure}

Within the scheme followed in Ref.~\cite{Benhar:2010nx} the occurrence
of 2p2h final states is described by the continuum part of the
spectral function, arising from nucleon-nucleon correlations,
and there, this contribution was found to be quite small.  The 2p2h contribution included  
in the spectral function corresponds only to mechanisms that can be cast as a nucleon selfenergy, as that depicted in the top panel of Fig.~\ref{fig:diag}. From electron-nucleus QE scattering studies, it is known 
that such contributions, though successful to describe the QE peak, can not account for the dip region, placed between the QE and the $\Delta$ peaks. 
In the case of neutrino scattering, since the energy of the incoming  beam is not fixed, the observed
energy of the outgoing charged lepton does not uniquely determine the energy transfer to the target, and hence the flux integration leads to collect contributions from different regimes, i.e. different reaction mechanisms, with about
the same probability (see the discussion of Fig.4 in Ref.~\cite{Benhar:2011ef}). In particular, mechanisms that populate the dip region lead to a considerable enhancement of the QE-like sample~\cite{Nieves:2011yp,Martini:2011wp}. 
A good starting point~\cite{Nieves:2011pp}
to evaluate these mechanisms is given by the set of  many body diagrams encoded in the bottom panel of Fig.~\ref{fig:diag}, constructed out of the elementary model for the $WN \to \pi N$ reaction derived in 
Refs.~\cite{Hernandez:2007qq,Hernandez:2010bx}\footnote{Note that the diagram showed in the top panel of  Fig.~\ref{fig:diag} is also implicit in the generic many body diagram depicted in the bottom 
panel, when the nucleon pole term, one of the seven  reaction mechanisms included in 
the model of Refs.~\cite{Hernandez:2007qq,Hernandez:2010bx}, is considered to account for the  two $WN \to \pi N$ transition  vertices (depicted  as shaded circles) that appear in the many body diagram.}.

\section{RPA, 2p2h and MiniBooNE 2D cross sections}

We will focus on neutrino cross sections, though the discussion runs in parallel for the case of anti-neutrino reactions~\cite{Nieves:2013fr, Martini:2013sha}. As mentioned in the introduction,  
the consideration of the 2p2h contributions allows to describe \cite{Nieves:2011yp,Martini:2011wp}  the MiniBooNE CCQE-like flux
averaged double differential cross section $d\sigma/dT_\mu/d\cos\theta_\mu$ \cite{AguilarArevalo:2010zc} with values of $M_A$ around 1 GeV. Thus, for instance the analysis of Ref.~\cite{Nieves:2011yp} finds  $M_A=1.077 \pm 0.027$ GeV
from a best fit  to the whole MiniBooNE data set, or $M_A = 1.007 \pm 0.034$ GeV, when  a transfer momentum threshold $q_{\rm cut}$ = 400 MeV is implemented, as suggested in \cite{Juszczak:2010ve}. This cut eliminates 14 of the
137 measured  bins that involve very low momenta, and for which a more detailed treatment of the nuclear degrees of freedom might be necessary. In both fits, only the axial mass 
$M_A$ and an overall normalization scale, $\lambda$, were adjusted to data. The obtained 
$\chi^2/dof$ turned out to be well below 0.5 and $\lambda \sim 0.9$, consistent with the global normalization
uncertainty of 10.7\% quoted in \cite{AguilarArevalo:2010zc}. 

\begin{figure}[tb] 
\begin{center}
\includegraphics[width=0.5\textwidth]{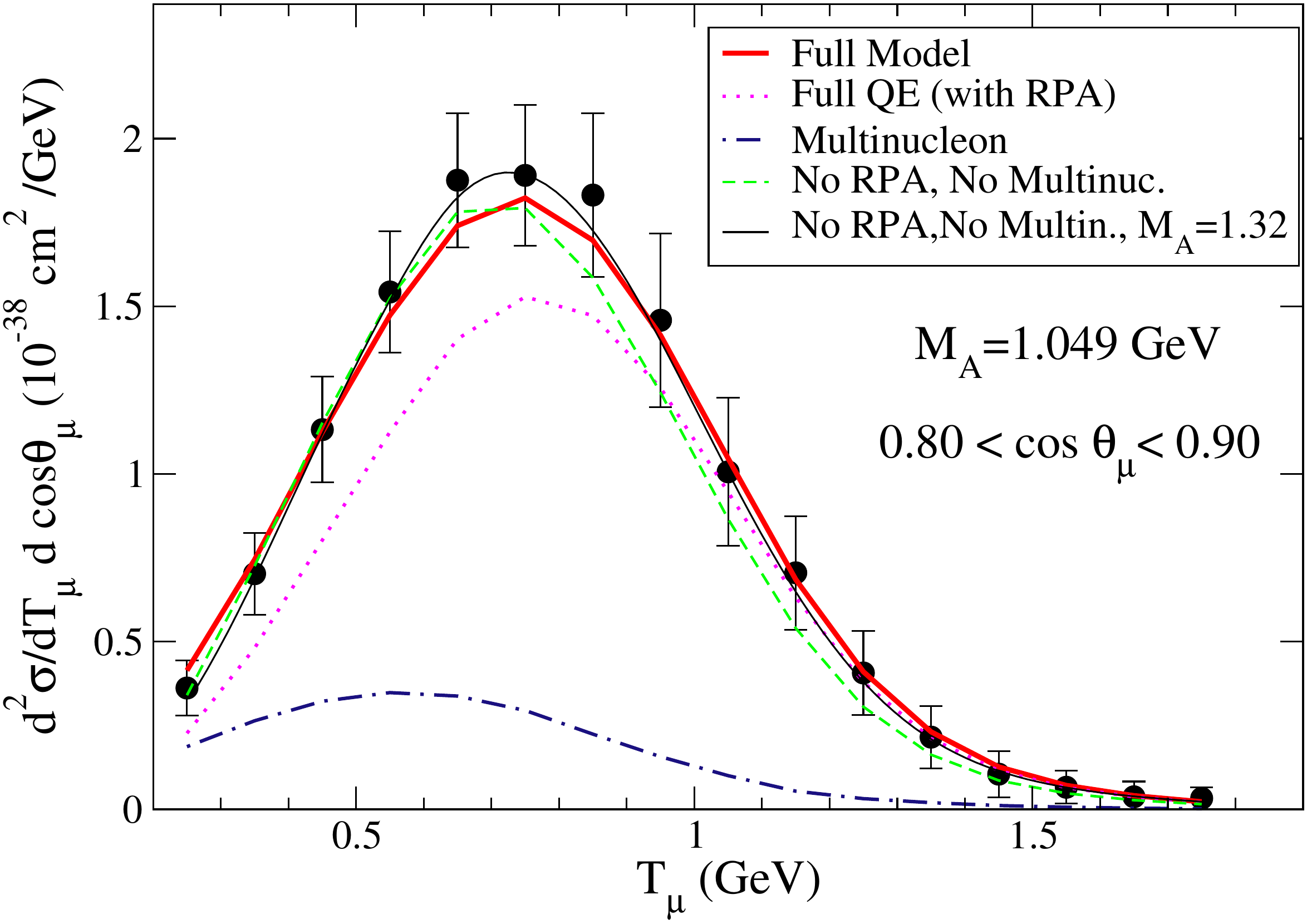}
\caption{ Muon angle and kinetic energy distribution $d\sigma/dT_\mu/d\cos\theta_\mu$ for the interval $0.80 < \cos \theta_\mu <
0.90$, per neutron. Experimental data from Ref.~\cite{AguilarArevalo:2010zc} and calculation with $M_A = 1.32$ GeV are multiplied by 0.9. The axial mass for the other curves is $M_A = 1.049$ GeV.}\label{fig:rpaMB}
\end{center}
\end{figure}

We would like to stress that, not only multinucleon mechanisms, but also RPA corrections turn out to be essential to determine axial masses consistent with the world average.  
Medium polarization or collective RPA correlations account for the change of the electroweak coupling strengths, from their free nucleon values, due to the presence of strongly interacting nucleons~\cite{Nieves:2004wx}. 
In Fig.~\ref{fig:rpaMB}, obtained within the model of Refs.~\cite{Nieves:2004wx} (QE) and  \cite{Nieves:2011pp,Nieves:2011yp} (2p2h), we see that RPA strongly decreases the cross section at low energies, while multinucleon mechanisms accumulate their contribution at
low muon energies and compensate for that depletion. Therefore, the final picture is that of a delicate balance between a dominant single nucleon scattering, corrected by collective effects,
and other mechanisms that involve directly two or more nucleons. Both effects can be mimicked by using a large $M_A$ value as done in the original experimental analysis~\cite{AguilarArevalo:2010zc}.  However, neglecting 
either of the two effects would lead to a poor description of the data.

\begin{figure}[tb] 
\begin{center}
\includegraphics[width=0.45\textwidth]{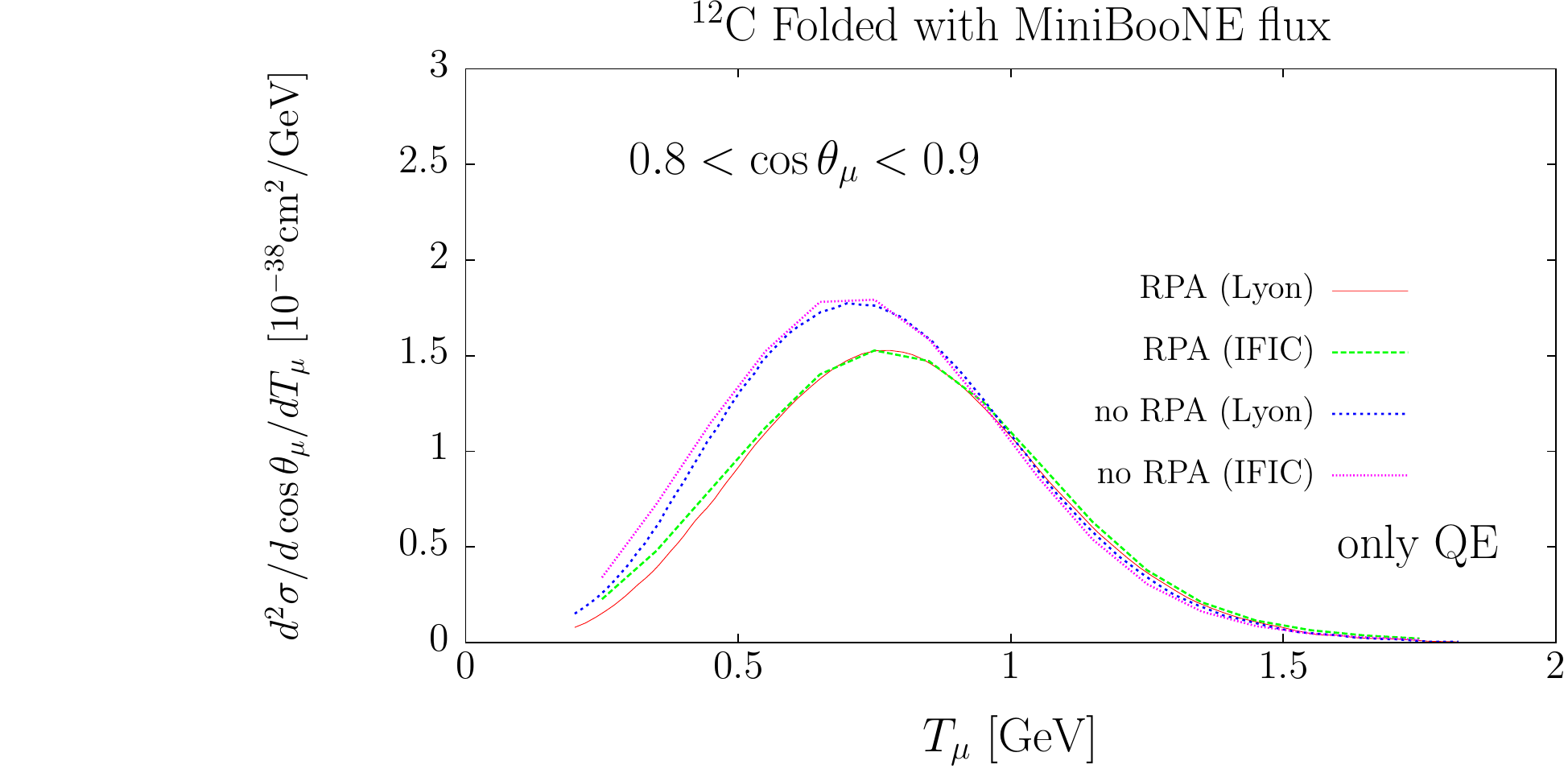}\\\vspace{0.5cm}
\includegraphics[width=0.45\textwidth]{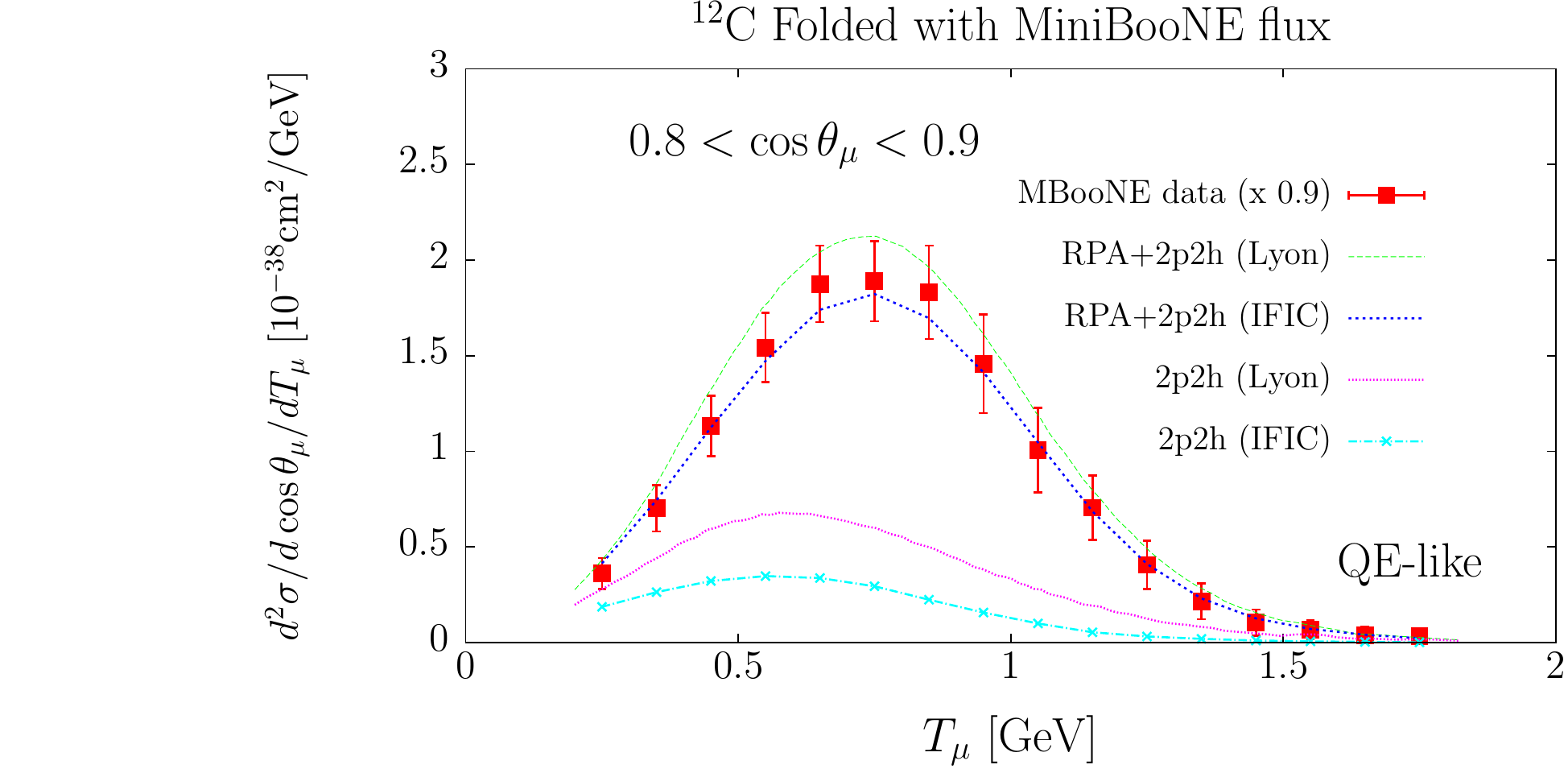}
\caption{Top: Predictions of the models of Refs.~\cite{Nieves:2004wx} (IFIC) and \cite{Martini:2009uj,Martini:2010ex, Martini:2011wp} (Lyon) for CCQE $\nu_\mu−^{12}$C double
differential cross section per neutron in the angular window $0.8 <  \cos \theta_\mu < 0.9$. The cross sections are calculated with a value of $M_A\sim 1$ GeV and averaged with the MiniBooNE flux. 
Results with and without RPA are shown. Bottom: 2p2h cross sections from 
the models of Refs.~\cite{Nieves:2011pp,Nieves:2011yp}  and \cite{Martini:2009uj,Martini:2010ex, Martini:2011wp} are compared and added to the RPA QE results. For comparison  experimental data from Ref.~\cite{AguilarArevalo:2010zc}, scaled down  by a factor 
0.9, are also displayed.
}\label{fig:IFIC-Lyon}
\end{center}
\end{figure}

M. Martini and collaborators find similar results \cite{Martini:2011wp}, since their model contains the same ingredients: RPA correlation effects and multinucleon mechanisms. As shown in the top panel of 
Fig.~\ref{fig:IFIC-Lyon}, the predictions of Ref.~\cite{Nieves:2004wx} for QE cross sections (labeled as IFIC in the figure), with and without RPA corrections, 
agree quite well with those obtained in \cite{Martini:2009uj,Martini:2010ex, Martini:2011wp} (labeled as Lyon in the figure). However, both  approaches differ in about a factor of two in their estimation of the
size of the multinucleon effects, as seen in the bottom panel of Fig.~\ref{fig:IFIC-Lyon}. As a consequence of this reduced 2p2h contribution, the IFIC predictions favor a global normalization scale, $\lambda$, of about 0.9~\cite{Nieves:2011yp}, 
which is not required by the Lyon model. As already mentioned, this  value of $\lambda$ is consistent with the MiniBooNE estimate of a total normalization error of 10.7\%. The evaluation in ~\cite{Nieves:2011pp,Nieves:2011yp} 
of multinucleon emission contributions to the cross section is fully microscopic and it starts from a state-of-the-art model~\cite{Hernandez:2007qq} for the $WN \to \pi N$ reaction at intermediate energies\footnote{
In addition to the $\Delta-$mechanism, the model includes also some background terms required by chiral symmetry. The dominant axial $N\Delta$  transition form factor is fitted to the flux-averaged 
$\nu_\mu p\to \mu^−p\pi^+$ ANL $q^2-$differential and BNL total cross section data, taken into account deuteron effects~\cite{Hernandez:2010bx}. The model was recently extended to higher energies above the $\Delta$ resonance region by adding 
a new resonant contribution corresponding to the $D_{13}(1520)$ nucleon excited state~\cite{Hernandez:2013jka}, which according to Ref.~\cite{Leitner:2008ue} and besides the $\Delta(1232)$, is the only resonance 
playing a significant role for neutrino energies below 2 GeV.} and contains terms, which were either not considered or only approximately
taken into account in  \cite{Martini:2009uj,Martini:2010ex, Martini:2011wp}. An example of such an approximation is the use
of a computation of the 2p2h mechanism for the $(e,e')$ inclusive reaction~\cite{Alberico:1983zg}
without modification to include axial-vector contributions, and their interference terms.

%

\section{Neutrino energy reconstruction}
Neutrino oscillation probabilities depend on the neutrino energy, unknown for broad fluxes
and often estimated from the measured angle and energy of the outgoing charged lepton. The specific reconstruction procedure is
determined by assuming QE kinematics for the event [$q^0 = -q^2/2M$, $q^\mu$ 
is the four-momentum of the $W$ gauge boson].  Neglecting binding energy and the difference between proton and neutron
masses, the estimate for the incident neutrino energy is 
\begin{equation}
E_{\rm rec} = \frac{M
  E_\mu-m_\mu^2/2}{M-E_\mu+|\vec{p}_\mu|\cos\theta_\mu}\label{eq:defereco}
\end{equation}
given the measured muon energy $E_\mu,$ three momentum $\vec{p}_\mu$  and the $W$ boson is
absorbed by a nucleon of mass $M$ at rest.

%
%

\begin{figure}[tb] 
\begin{center}
\includegraphics[width=0.5\textwidth]{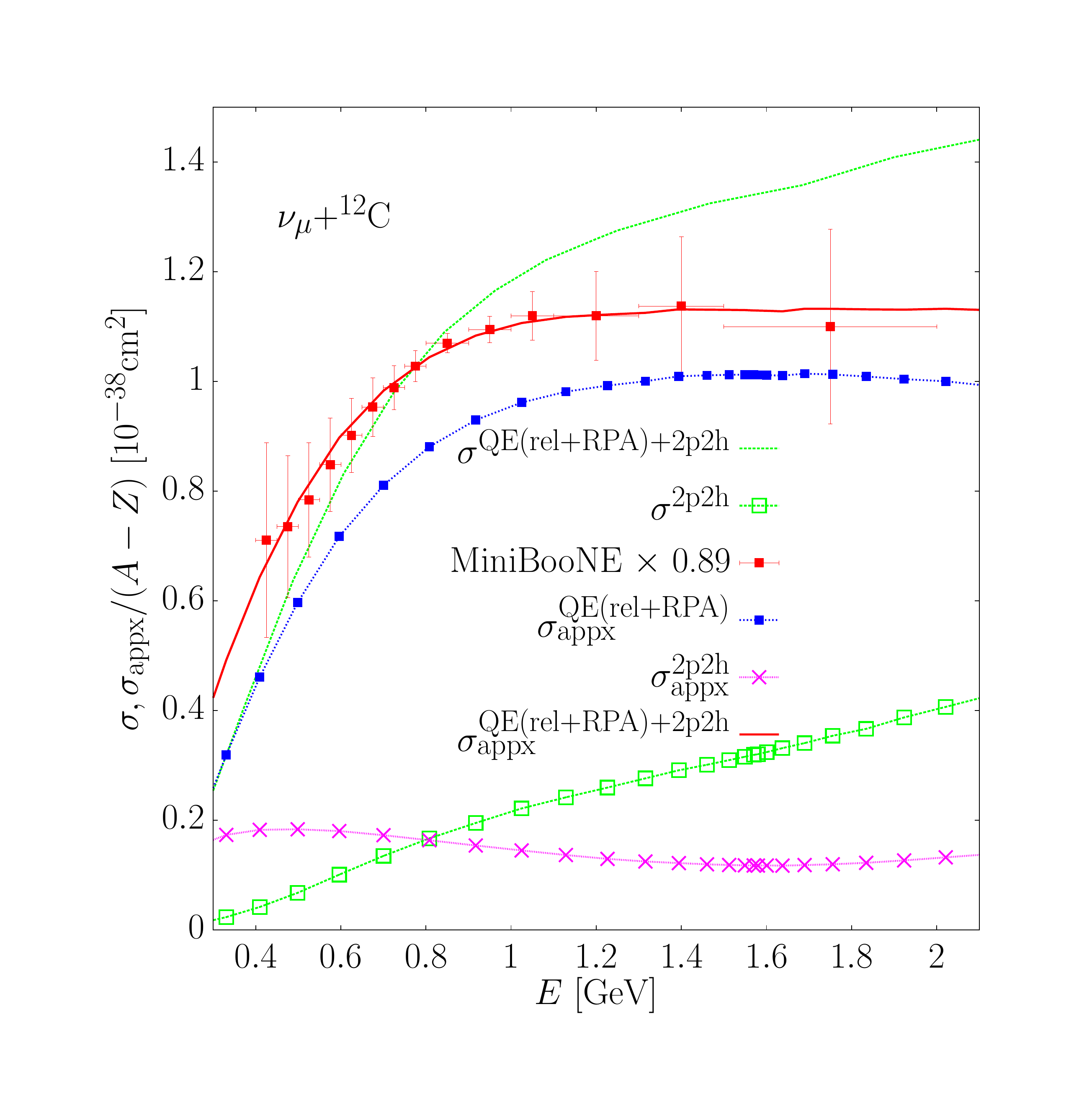}
\caption{Theoretical ($\sigma$) and approximate ($\sigma_{\rm appx}$) CCQE-like integrated cross sections obtained in \cite{Nieves:2012yz} as a function of the true neutrino energy. The MiniBooNE data \cite{AguilarArevalo:2010zc}
and errors (shape) have been re-scaled by a factor 0.89. All theoretical results have been obtained with the model of Refs. ~\cite{Nieves:2011pp,Nieves:2011yp,Nieves:2004wx} and $M_A = 1.05$ GeV.
}\label{fig:ereco}
\end{center}
\end{figure}

For each value of the reconstructed neutrino energy, there exists a distribution of true neutrino energies that 
give rise to events whose muon kinematics would lead to the given value of $E_{\rm rec}$. In the case of genuine QE events,
this distribution is  peaked around the true neutrino energy to make the algorithm in Eq.~(\ref{eq:defereco})
sufficiently accurate for most purposes~\cite{Nieves:2012yz,Martini:2012fa,Lalakulich:2012hs}. It has long been known that 
the background from $\Delta$ production
in a QE-like sample is reconstructed with anomalously low energy and low
$Q^2(=-q^2)$ when using Eq.~(\ref{eq:defereco}), and is accounted for using a model of the
$\Delta$ background.  This is also true for the 2p2h, which for a
real flux has a long tail of true energies associated with each $E_{\rm rec}$.  In a
broad, peaked flux spectrum, this makes the approximation of Eq.~(\ref{eq:defereco}) unreliable \cite{Nieves:2012yz,Martini:2012fa}, since the redistribution of strength from high to low energies gives
rise to a sizable excess (deficit) of low (high) energy neutrinos.
This is illustrated in Fig.~\ref{fig:ereco} for the flux unfolded CCQE-like cross section reported by the MiniBooNE 
Collaboration~\cite{AguilarArevalo:2010zc}. There,  different predictions taken from Ref.~\cite{Nieves:2012yz}, 
together with the data are shown. The unfolding
procedure used in  \cite{AguilarArevalo:2010zc} does not appreciably distort the genuine QE events, however the situation
is drastically different for the 2p2h contribution, where a systematic and significant distortion of its
energy shape is produced. This  systematic effect certainly increases the uncertainty on the extracted oscillation signal, and points out the impossibility to extract cross sections as a function of the 
neutrino energy in a model independent manner. These conclusions were corroborated within the model of Refs.~\cite{Martini:2009uj,Martini:2010ex, Martini:2011wp} in their  later work of 
Ref.~\cite{Martini:2012uc}. 

On the other hand, it is remarkable the agreement exhibited in Fig.~\ref{fig:ereco} between the MiniBooNE pseudo-data shape and the predictions of the model 
derived in Refs.~\cite{Nieves:2011pp, Nieves:2011yp,Nieves:2004wx}, when the approximate unfolding procedure used in \cite{AguilarArevalo:2010zc} was followed. The agreement, though also quite good, is not as good when the model of 
Refs.~\cite{Martini:2009uj,Martini:2010ex, Martini:2011wp} is used instead (see Fig. 14 of Ref.~\cite{Martini:2012uc}). Note however, that the latter model provides a 
better description of the data than that obtained within the IFIC model, when  
 the corrections induced by the unfolding procedure are not taken into account (this can be seen in the left panel of Fig.18 in Ref.~\cite{Nieves:2011pp} or by comparing the solid green solid line in Fig.~\ref{fig:ereco} 
with the red solid line of the above mentioned Fig. 14 of Ref.~\cite{Martini:2012uc}).

\section{Results at higher energies}

\begin{figure}[tb]
\begin{center}
\includegraphics[width=0.25\textwidth]{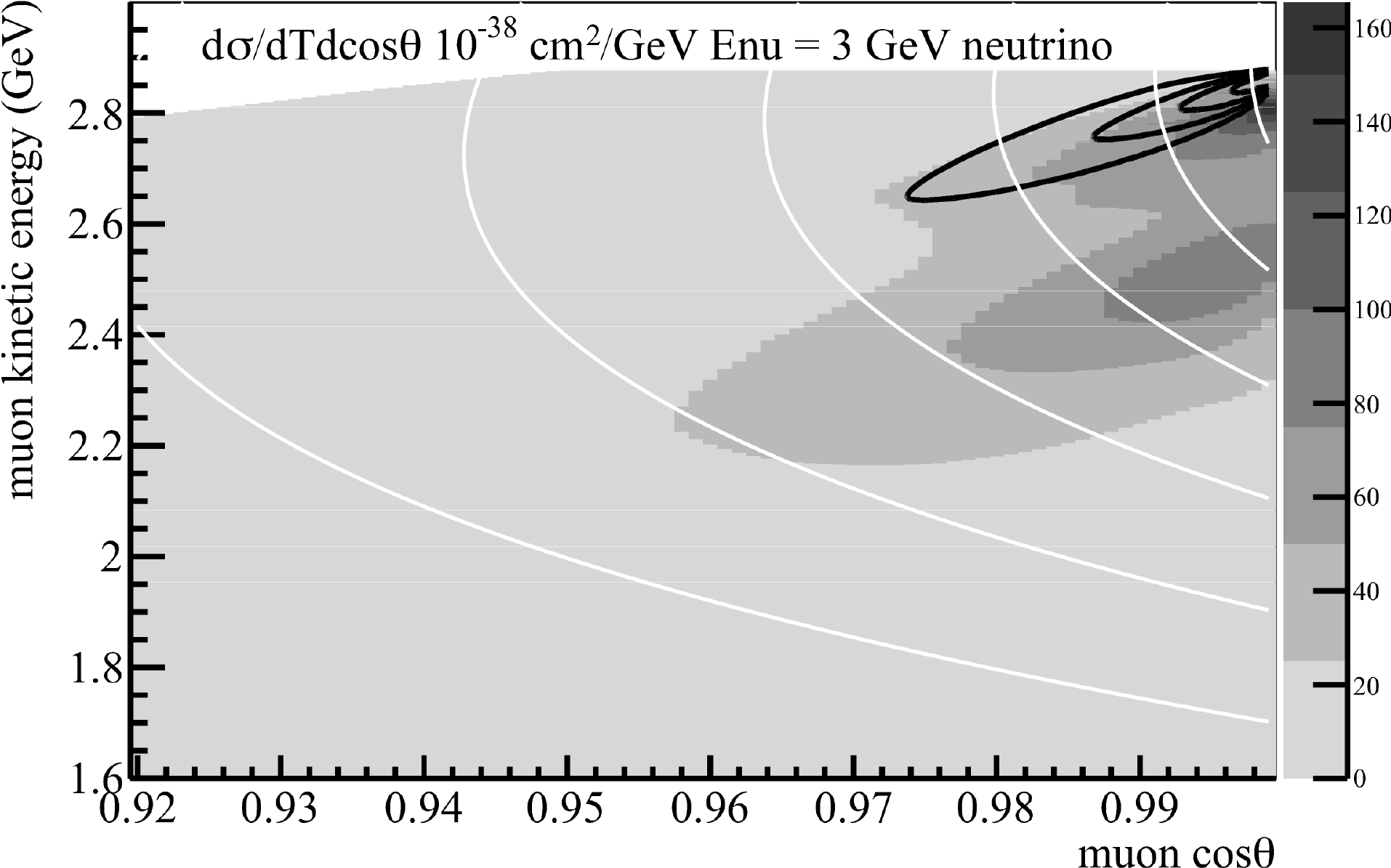}\\\vspace{0.5cm}
\includegraphics[width=0.25\textwidth]{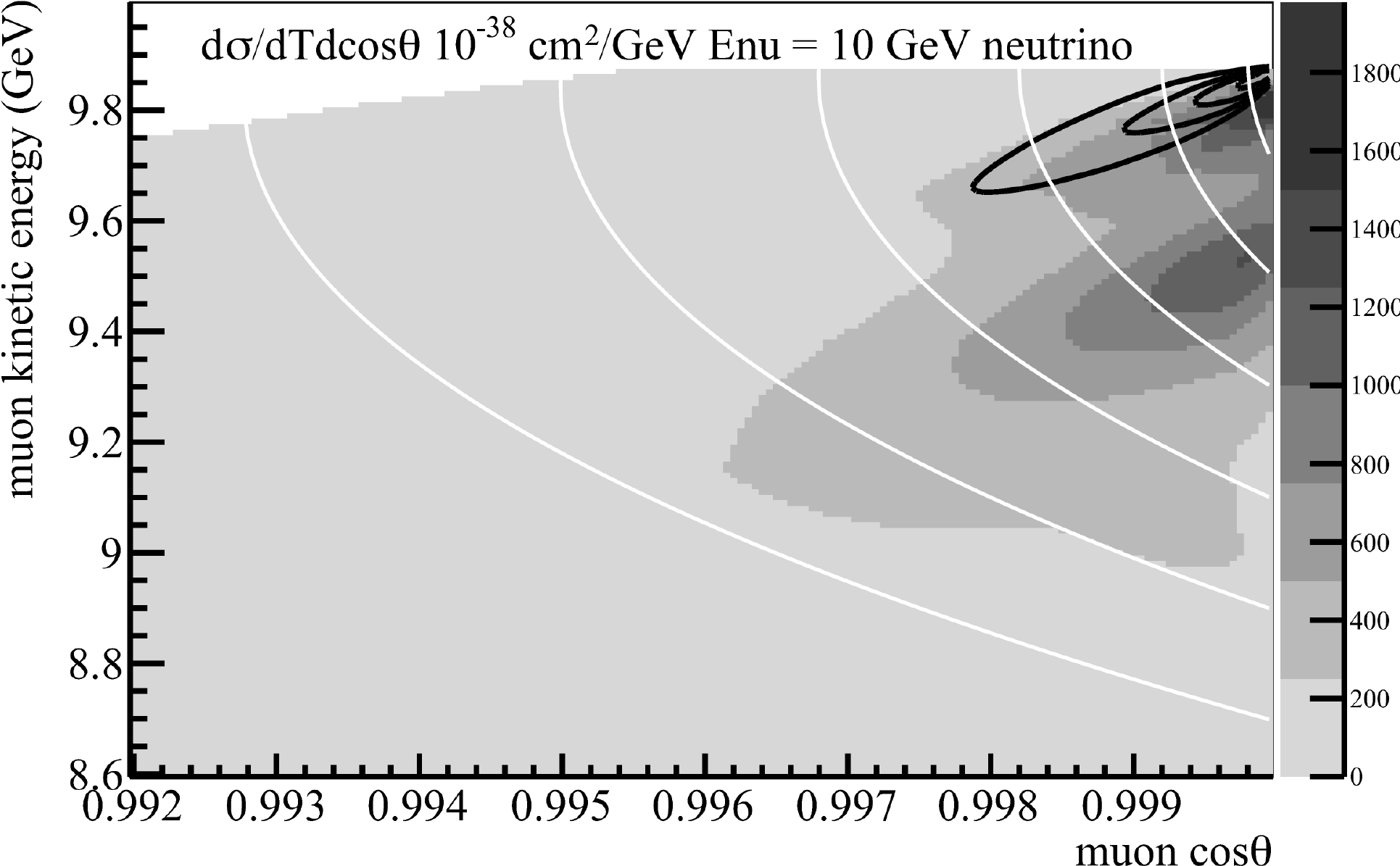}
\end{center}
\caption{\label{fig:3} Double differential 2p2h cross section for
  neutrino-carbon interactions at energies of 3 and 10 GeV. The
  black contours show the location of the genuine QE events, while the
  white ones show lines of constant three-momentum transfer from 0.2
  to 1.2 GeV (see Ref.~\cite{Gran:2013kda} for further details).}
\end{figure}
\begin{figure}[tb] 
\begin{center}
\includegraphics[width=0.4\textwidth]{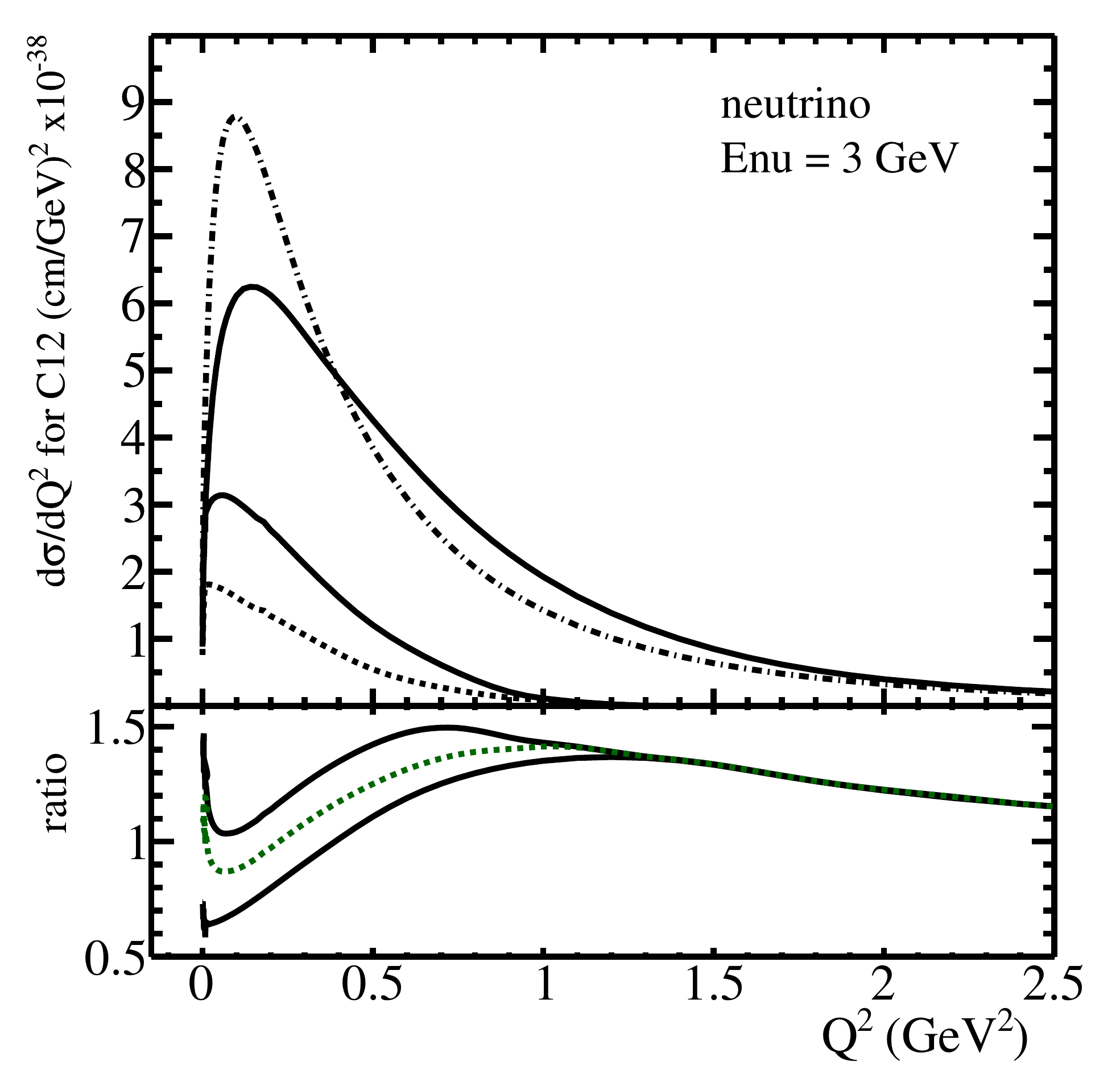}
\caption{The $Q^2(=-q^2)$ differential cross section for QE (solid upper curve) and 2p2h (solid lower curve)  contributions,
for a neutrino incoming energy of 3 GeV. The upper dashed line stands for  the QE
without RPA (similar to the standard treatment for
neutrino experiments), while the lower dashed line is the 2p2h cross section
without the $\Delta$ absorption component, 2p2h$_{{\rm no}\Delta}$ (see Ref.~\cite{Gran:2013kda} for details). The lower solid ratio line is QE$_{\rm RPA}$/QE$_{\rm noRPA}$, the dashed ratio line is (QE$_{\rm RPA}$+ 
2p2h$_{{\rm no}\Delta})$/QE$_{\rm noRPA}$, and the upper ratio line is (QE$_{\rm RPA}$+ 
2p2h$_{{\rm with}\Delta})$/QE$_{\rm noRPA}$. In all cases, the QE lines are the
complete cross section calculated with the model of Ref.~\cite{Nieves:2004wx}, whereas  the 2p2h lines, calculated with the model of  Refs.~\cite{Nieves:2011pp,Nieves:2011yp}, 
truncate the integration at $|\vec{q}\,| < 1.2$ GeV.}\label{fig:QERPA}
\end{center}
\end{figure}

We have extended to higher energies the results from the microscopic model of Refs.~\cite{Nieves:2004wx} (QE) and  \cite{Nieves:2011pp,Nieves:2011yp} (2p2h), both for neutrino and anti-neutrino CC reactions.
Limiting the calculation to three momentum transfers less than 1.2 GeV, we find~\cite{Gran:2013kda} as the neutrino energy increases, up to 10 GeV, the 2p2h contribution
saturates to $\sim 30$\% of the QE cross section (see Fig.~\ref{fig:3}). In principle, there is no reason for this trend to change
drastically at even higher energies. This brings us to a question that remains open: the
compatibility of the MiniBooNE results with the NOMAD one, $M_A = 1.05 \pm 0.02~({\rm stat})~ \pm 0.06~ {\rm (syst)}$ GeV, quoted above. The answer is not obvious and requires
further investigations.  The NOMAD experiment analyzed a set of QE-like
interactions on carbon \cite{Lyubushkin:2008pe} whose flux has an average
energy of 25.9 GeV for neutrino and 17.6 GeV for anti-neutrino.  This experiment includes  two-track sample events, which is primarily
$Q^2$ above 0.3 GeV$^2$.  Events from 2p2h production should be especially rejected, and also
QE events where the outgoing hadron rescattered as it exited the nucleus,
by the requirement of high momentum transfer and a proton matching the CCQE
hypothesis;  it should be an especially
pure sample of QE kinematics.
These 2p2h and rescattered QE events are either in the NOMAD
one-track sample or are rejected two-track events and not considered in the
analysis at all.

It is worth nothing the relative deficit in the data
at $Q^2 = 0.3$ GeV$^2$ and excess at 1.5 GeV$^2$, compared to their QE model without
RPA (see Fig.14 of Ref.~\cite{Lyubushkin:2008pe}).  Their fit to the shape of this distribution
apparently balances this against the lowest $Q^2$ data points.  The former
behavior has some resemblance to the findings of Ref.~\cite{Gran:2013kda}, in particular
with the 
lower solid ratio line showed in Fig.~\ref{fig:QERPA} that stands for  QE$_{\rm RPA}$/QE$_{\rm noRPA}$ calculated with 
the model of Ref.~\cite{Nieves:2004wx}. 

Taking into account the RPA series leads to a
large $Q^2-$shape distortion, with the 2p2h component filling in the suppression
at very low $Q^2$, as commented before and also shown in Fig.~\ref{fig:QERPA}.    The low $Q^2$ suppression is a combination of both
short and long range correlation effects. The trend moving toward $Q^2= 1.1$ GeV$^2$ is an enhancement of the cross
section but leaves the region where the model of Ref.~\cite{Nieves:2004wx} was tuned
to other low energy nuclear data. The in-medium effective $NN$ interaction used to compute the RPA
correlations is not realistic at high three momentum and energy transfers,
and thus the model suffers from larger uncertainties. 
However, a model independent prediction is that the RPA corrections should disappear (ratio goes to 1.0) at very large $Q^2$ values, because this is a collective effect which strength decreases when
sizes larger than one nucleon are no longer being probed. Hence in any realistic model, one should expect a qualitative $Q^2$ behaviour similar to that exhibited by the QE$_{\rm RPA}$/QE$_{\rm noRPA}$ ratio line depicted in  Fig.~\ref{fig:QERPA}: 
low $Q^2$ suppression, followed by an enhancement that could even give rise to a net increase of the cross section, and finally all RPA effects should disappear for sufficiently high $Q^2$ values. The most robust predictions 
of the model of Ref.~\cite{Nieves:2004wx} are those related to the RPA diminution of events in the low $Q^2$ region, since the correlation effects in this model
are tuned to low energy nuclear phenomena, such as pion and electron scattering and muon capture on nuclei, where they are essential for a good description of data. Besides in Fig.~\ref{fig:QERPA}, the effects of 
the  $\Delta$ component in the 2p2h contribution can be also seen (see details in \cite{Gran:2013kda}). This is an important issue, since a portion of the cross section involving  $\Delta$ absorption, might be incorporated
into modern event generators via the treatment of $\Delta$ and/or pion final state re-interactions in the nucleus.

\begin{figure}[tb] 
\begin{center}
\includegraphics[width=0.4\textwidth]{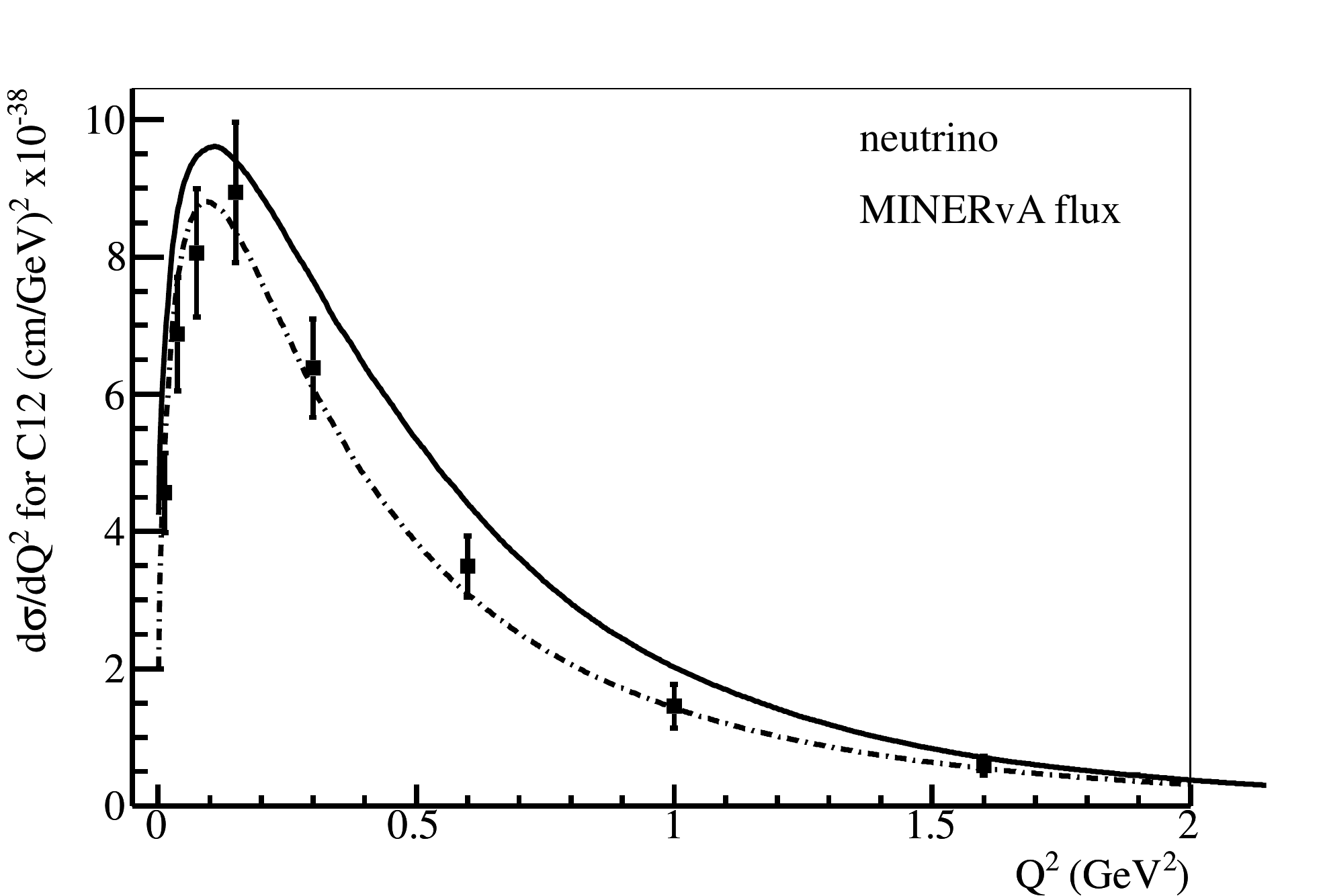}\\\vspace{0.3cm}
\includegraphics[width=0.4\textwidth]{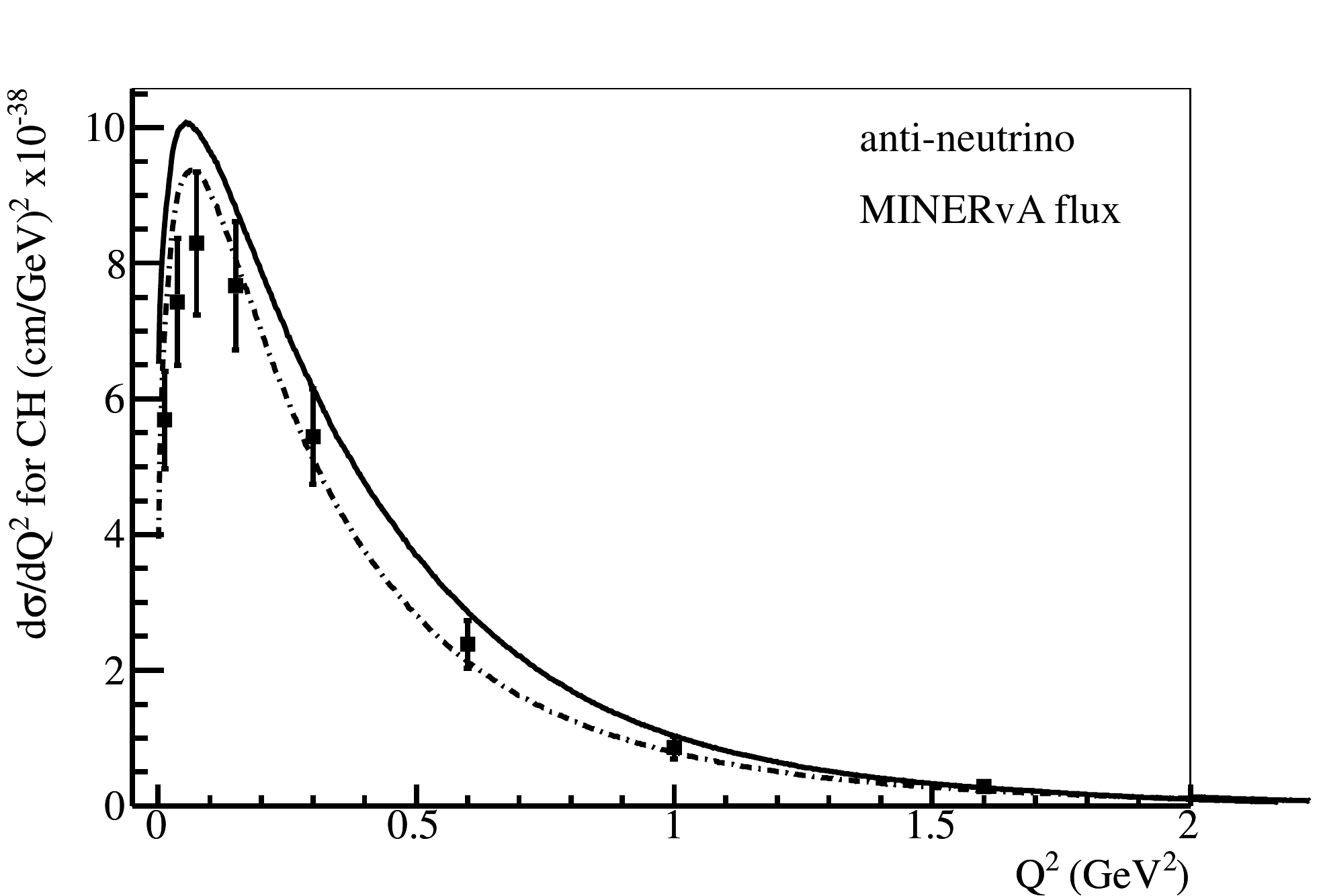}
\caption{Differential $Q^2$ distribution averaged over the MINERvA $\nu_\mu$ (top) and $\bar\nu_\mu$ (bottom) fluxes \cite{Fiorentini:2013ezn,Fields:2013zhk} as a function of the reconstructed $Q^2$.
Solid lines stand for results with 2p2h and QE with RPA effects~\cite{Nieves:2011pp,Nieves:2011yp,Nieves:2004wx}, while dot-dashed lines stand for results without RPA and without 2p2h effects~\cite{Nieves:2004wx}. 
Data are from \cite{Fiorentini:2013ezn,Fields:2013zhk}. The 2p2h cross sections truncate the integration at $|\vec{q}\,| < 1.2$ GeV.}\label{fig:MINERVA}
\end{center}
\end{figure}

The predictions of the model derived in \cite{Nieves:2011pp,Nieves:2011yp,Nieves:2004wx},
averaged over the neutrino and anti-neutrino fluxes at MINERvA were compared to 
data in \cite{Gran:2013kda}  for the reconstructed $q^2-$distribution obtained using the anti-neutrino/neutrino reconstructed energy. As can be seen in Fig.~\ref{fig:MINERVA},
the agreement is quite good, with a slight overestimation of the data. The impact of the
reconstruction procedure in the case of MINERvA flux is small. The model has the qualitative features and
magnitude to provide a reasonable description of the data.

%

\section{Acknowledgements}
This work has been produced with the support of the Spanish
Ministerio de Econom\'\i a y Competitividad and European FEDER funds
under the contracts FIS2011-28853-C02-01, FIS2011-28853-C02-02, FIS2011-24149,
FPA2011-29823-C02-02, CSD2007-0042 and SEV-2012-0234, the Generalitat
Valenciana and the Junta de Andaluc\'\i a under contracts  PROMETEOII/2014/0068 and FQM-225, and the U.S. National
Science Foundation under Grant Nos. 0970111 and 1306944.





\bibliography{neutrinos}







\end{document}